# Motivations to modify special relativity

Jian-Miin Liu*
Department of Physics, Nanjing University
Nanjing, The People's Republic of China
*On leave. E-mail address: liu@phys.uri.edu

ABSTRACT

In the framework of special relativity, all particles are point-like or string-like. This nature of particles has caused the divergence difficulties in quantum field, string and superstring theories. In the framework of special relativity, due to the non-uniformity of the μ-space and phase space in the usual inertial coordinate system, Boltzmann's hypothesis of the equality of the probability of equal volume element is no longer appropriate. That makes it very difficult to construct Lorentz-invariant statistical mechanics and thermodynamics for many-particle systems. Besides, some observations on special relativity itself and its experimental facts are also reported. The conclusions from these observations are: Special relativity is not an ultimate theory; Some modification is needed; Any modification must not violate the constancy of the light speed and the local Lorentz invariance; It seems that we have to change the assumption on local structures of gravity-free space and time in special relativity.

## I. OBSERVATION 1: SPECIAL RELATIVITY AND QUANTUM MECHANICS

Special relativity constitutes the speed of light as a speed limit for the transport of matter or energy and the transmission of information or causal connection. Any "faster than c" process between two events at distant points in space, including instantaneous quantum connection as it appears to be [1-3], is not allowed. In the framework of special relativity, all particles are point-like. Otherwise, it is hard to understand how a finite-sized particle as a whole is set in motion when a force acts on it at its edge, and it is hard to carry out Lorentz-invariant calculations for such finite-sized particles.

Point-like nature of particles has caused the divergence difficulties in classical field theory, as well as in quantum field theory. Indeed, the infinite self-energy of a sizeless electron in quantum electrodynamics was known as early as 1929 [4], while it was known in classical electrodynamics even earlier. Phenomenological substituting several finite experimental values of particle masses and charge for their infinities in theoretical calculations, physicists developed renormalization techniques to remove all divergence in some quantized field systems. However, not all quantized field systems are renormalizable, and it is difficult to accept such a kind of renormalizability as a basic physical principle to truncate non-renormalizable systems. Feynman said: "renormalization of a quantity gives up any possibility of calculating that quantity" [5]. The renormalized quantum field theory fails to explain a class of important phenomena, mass differences in the groups of neutron-proton, the π-mesons, the K-mesons, the Σ-baryons and the Ξ-baryons.

Physicists created the concepts of one-dimensional extended objects, called strings. Although the concepts of strings are not contrary to special relativity, string or superstring theories do not get rid of the divergence difficulties [6]. String or superstring theories still rely on renormalizability to eliminate divergence.

Quantum field theory is a theory founded on the unification of special relativity and quantum mechanics. As it is the case that both classical field theory and quantum field theory are plagued by the divergence difficulties, the origin of the divergence difficulties seems to lie in special relativity.

## II. OBSERVATION 2: SPECIAL RELATIVITY AND STATISTICAL MECHANICS

Special relativity is a part of the laws of Nature. It is natural to try to move statistical mechanics from its pre-relativistic mechanics base to relativistic mechanics base. This is why the continuous efforts in constructing Lorentz-invariant statistical mechanics and thermodynamics for many-particle systems began so soon after the birth of special relativity [7-16], though all these efforts ended in failure. So far, we have not had an acceptable Lorentz-invariant statistical mechanics and thermodynamics for many-particle systems.

Starting with the four-dimensional Minkowskian space,
$$d\Sigma^2 = \delta_{ij} dx^i dx^j, \quad i,j=1,2,3,4, \tag{1}$$
where $x^1=x$, $x^2=y$, $x^3=z$ and $x^4=ict$, c is the speed of light, some authors [11-16] found the four-dimensional tangent space which was further identified with the momentum space,
$$P^2 = \delta_{ij} p^i p^j, \quad i,j=1,2,3,4, \tag{2}$$
where $p^i = m dx^i/dt'$, $dt' = (1-y^2/c^2)^{1/2} dt$. Having the Minkowskian space and momentum space, they proposed an eight-dimensional μ-space and an 8N-dimensional phase space with constrain,
$$\delta_{ij} p^i p^j = -m^2 c^2 \tag{3}$$
for each particle. The μ-space and phase space are uniform. Their volume elements
$$d^4x = dx^1 dx^2 dx^3 dx^4 \tag{4a}$$
and
$$d^4p = dp^1 dp^2 dp^3 dp^4 \tag{4b}$$
are invariant under the Lorentz transformation.

The uniformity of the μ-space and phase space convinces them of Boltzmann's hypothesis of the equality of the probability of equal volume element and, hence, of the Boltzmann-like distribution in the μ-space and the Gibbs-like microcanonical and canonical (after Gibbs's hypothesis) distributions in the phase space. The invariance of $d^4x$ and $d^4p$ specifies the invariant properties of distribution functions in these distributions under the Lorentz transformation. To determine the mentioned Boltzmann-like and Gibbs-like distribution functions, they wanted to find out suitable energy formulas which are invariant.

However, no such energy formulas had been found. The energy of an individual particle is not an invariant. It is transformed as a component of the momentum-energy vector. Total energy of a many-particle system is a sum of kinetic energies and interaction energies of its all particles and particle pairs, at an instant, in the many-particle system. As simultaneity at distant space points in a given inertial frame of reference is no longer simultaneous in any different inertial frame of reference, the transformation properties of the total energy of the many-particle system are quite indefinite.

Actually, according to Rund [33], the tangent space at a point on a Riemannian manifold is an affine space centered at the point. In the case of the Minkowskian manifold, the tangent space coincides with the manifold itself. But this tangent space is not necessarily identical with the momentum space. In special relativity, the velocity space is [28-32],
$$dY^2 = H_{rs}(y) dy^r dy^s, \quad r,s=1,2,3, \tag{5a}$$
$$H_{rs}(y) = c^2 \delta^{rs}/(c^2-y^2) + c^2 y^r y^s/(c^2-y^2)^2, \quad \text{real } y^r \text{ and } y<c, \tag{5b}$$
in the usual velocity-coordinates $\{y^r\}$, r=1,2,3, where $y^r = dx^r/dt$ is the well-defined Newtonian velocity, $y = (y^r y^r)^{1/2}$. This velocity space is non-uniform. As a result, the four-dimensional momentum space can not be uniform, and the μ-space and phase space can not be uniform, either. Boltzmann's hypothesis of the equality of the probability of equal volume element is no longer appropriate in the framework of special relativity. Therefore, we have neither the Boltzmann-like nor the Gibbs-like distributions in the framework of special relativity. It is very tough, in the framework of special relativity, to construct Lorentz-invariant statistical mechanics and thermodynamics for many-particle systems.

### III. OBSERVATION 3: EXPERIMENTAL FACTS

Because of technological limitations, in the earlier experiments testing the constancy or isotropy of the speed of light, light was propagated in a closed path. The favorite conclusions from these experiments are obviously for the constancy of the speed of the round-trip light, not for that of the one-way light. As a result of technological advances, many experiments have been performed in the manner that light propagates in a one-way. Two research groups, of Turner and Hill [17], and of Champeney et al [18], placed a $Co^{57}$ source near the rim of a standard centrifuge with an iron absorber near the axis of rotation. They used the Mossbauer effect to look for any velocity dependence of the frequency of the 14.4 KeV γ-rays as seen by the $Fe^{57}$ in the absorber. They established limits of $\Delta c/c < 2 \times 10^{-10}$ for the anisotropy in the one-way speed of light. Riis and his colleagues [19] compared the frequency of a two-photon transition in a fast atomic beam to that of a stationary absorber while the direction of the fast beam is rotated relative to the fixed stars and found the upper limit $\Delta c/c < 3.5 \times 10^{-9}$ firstly and $\Delta c/c < 2 \times 10^{-11}$ later for the anisotropy. The experiment of Krisher et al [20] was made using highly stable hydrogen-maser frequency standards (clocks) separated by over 21 km and connected by a ultrastable fiber optics link. The limits yielded from the experimental data are respectively $\Delta c/c < 2 \times 10^{-7}$ for linear dependency and $\Delta c/c < 2 \times 10^{-8}$ for quadratic dependency on the velocity of the Earth with respect to the cosmic microwave background.

All experimental tests of the constancy of the speed of one-way light can also be interpreted as the tests of the local Lorentz invariance. Nevertheless, since local Lorentz non-invariance implies a departure from the Einstein time dilation and an existence of preferred inertial frame of reference, the experiments done by McGowan et al [21], Bailey et al [22], Kaivola et al [23], Prestage et al [24], and Krisher et al [20] can be accounted as immediate testing of the local Lorentz invariance. Bailey et al, Kaivola et al, and McGowan et al verified the Einstein time dilation to an accuracy of $1 \times 10^{-3}$, $4 \times 10^{-5}$ and $2.3 \times 10^{-6}$ respectively. The experiments of Prestage et al and Krisher et al are sensitive to the effects of motion of their experimental apparatus relative to a preferred inertial frame of reference and sensitive to the form of time dilation coefficient of the hydrogen-maser clocks in use. The null results were produced in these two experiments for breakdown of the local Lorentz invariance.

Experiments clearly indicate the constancy of the speed of light, the Einstein velocity addition law and the local Lorentz invariance.

IV. OBSERVATION 4: SPECIAL RELATIVITY ITSELF

Einstein published his special theory of relativity in 1905 [25]. He derived the Lorentz transformation between any two usual inertial coordinate systems, which is the kinematical background for the physical principle of the Lorentz invariance. Two fundamental postulates stated by Einstein as the basis for his theory are (i) the principle of relativity and (ii) the constancy of the speed of light in all inertial frames of reference.

Conceptually, the principle of relativity implies that there exists a class of equivalent inertial frames of reference, any one of which moves with a non-zero constant velocity relative to any other one. Einstein wrote: "in a given inertial frame of reference the coordinates mean the results of certain measurements with rigid (motionless) rods, a clock at rest relative to the inertial frame of reference defines a local time, and the local time at all points of space, indicated by synchronized clocks and taken together, give the time of this inertial frame of reference."[26]. As defined by Einstein, each of the inertial frames of reference is supplied with motionless, rigid unit rods of equal length and motionless, synchronized clocks of equal running rate. Then, in each inertial frame of reference, an observer can employ his own motionless-rigid rods and motionless-synchronized clocks in the so-called "motionless-rigid rod and motionless-synchronized clock" measurement method to measure space and time intervals. By using this "motionless-rigid rod and motionless-synchronized clock" measurement method, the observer can set up his own usual inertial coordinate system, $\{x^r, t\}$, $r=1,2,3$, $x^1=x$, $x^2=y$, $x^3=z$. Postulate (ii) means that the speed of light is the same constant c in every such usual inertial coordinate system.

Besides two postulates (i) and (ii), special relativity also uses another assumption. This other assumption concerns the Euclidean structure of gravity-free space and the homogeneity of gravity-free time in the usual inertial coordinate system,

$dX^2 = \delta_{rs} dx^r dx^s$, $r,s=1,2,3$, (6a)
$dT^2 = dt^2$, (6b)

everywhere and every time.

The assumption Eqs.(6a-6b) and postulates (i) and (ii) together yield the Lorentz transformation between any two usual inertial coordinate systems. Indeed, though the assumption Eqs.(6a-6b) was not explicitly articulated, evidently having been considered self-evident, Einstein said in 1907: "Since the propagation velocity of light in empty space is c with respect to both reference systems, the two equations, $x_1^2 + y_1^2 + z_1^2 - c^2 t_1^2 = 0$ and $x_2^2 + y_2^2 + z_2^2 - c^2 t_2^2 = 0$, must be equivalent." [7]. Leaving aside a discussion of whether postulate (i) implies the linearity of transformation between any two usual inertial coordinate systems and the reciprocity of relative velocities between any two usual inertial coordinate systems, we know that the two equivalent equations, the linearity of transformation and the reciprocity of relative velocities lead to the Lorentz transformation.

Some physicists explicitly articulated the assumption Eqs.(6a-6b) in their works on the topic. Pauli wrote: "This also implies the validity of Euclidean geometry and the homogeneous nature of space and time." [28], Fock said: "The logical foundation of these methods is, in principle, the hypothesis that Euclidean geometry is applicable to real physical space together with further assumptions, viz. that rigid bodies exist and that light travels in straight lines." [29].

The Minkowskian space Eq.(1) is a four-dimensional version of the assumption Eqs.(6a-6b).

From the assumption Eqs.(6a-6b), one can find

$$Y^2 = \delta_{rs} y^r y^s, \tag{7}$$

where $Y = dX/dT$ is the velocity-length, and $y^r = dx^r/dt$, $r=1,2,3$, as said, is the Newtonian velocity. Eq.(7) embodies the velocity space,

$$dY^2 = \delta_{rs} dy^r dy^s. \tag{8}$$

This velocity space is characterized by boundlessness and the Galilean velocity addition law. On the other hand, special relativity owns the Lorentz transformation between any two usual inertial coordinate systems. That actually implies a finite velocity boundary at c and the Einstein law governing velocity additions, in other words, it is the velocity space defined in Eqs.(5a-5b) that stands in special relativity. We see an inconsistency.

## V. CONCLUSIONS

From observations 1 and 2: Special relativity is not an ultimate theory; Some modification is needed.

From observation 3: Any modification of special relativity must not violate the constancy of the light speed and the local Lorentz invariance.

From observation 4: It seems that we have to change the assumption Eqs.(6a-6b) in special relativity.

The generalized Finslerian structures of gravity-free space and time in the usual inertial coordinate system has been logically proposed [30-32].

ACKNOWLEDGMENT

The author greatly appreciates the teachings of Prof. Wo-Te Shen. The author thanks Prof. M. S. El Naschie for his supports of this work.